\documentclass[%
superscriptaddress,
amsmath,amssymb,
aps,
]{revtex4-2}

\usepackage{physics}
\usepackage{amsmath}
\usepackage{amsfonts}
\usepackage{caption}
\usepackage{graphicx}
\usepackage{upgreek}
\usepackage{bm}
\usepackage{siunitx}
\newcommand{\loss}[1]{\mathcal{L}_{\mathrm{#1}}}

\newcommand{\set}[1]{\left\{#1\right\}}
\newcommand{\psig}{\ket{\Psi_{0,0}^G}}
\newcommand{\copropagatingbasis}{\{\ket{\Psi_{n_x, n_y}^{C}}\}}

\begin{document}

\title{Mode mixing and losses in misaligned microcavities}

\author{William J Hughes}
\email{Corresponding Author: william.hughes@physics.ox.ac.uk}
\affiliation{Department of Physics, University of Oxford, Clarendon Laboratory, Parks Road, Oxford OX1 3PU, United Kingdom}
\author{Thomas H Doherty}
\affiliation{Department of Physics, University of Oxford, Clarendon Laboratory, Parks Road, Oxford OX1 3PU, United Kingdom}
\author{Jacob A Blackmore}
\affiliation{Department of Physics, University of Oxford, Clarendon Laboratory, Parks Road, Oxford OX1 3PU, United Kingdom}
\author{Peter Horak}
\affiliation{Optoelectronics Research Centre, University of Southampton, Southampton SO17 1BJ, UK}
\author{Joseph F Goodwin}
\affiliation{Department of Physics, University of Oxford, Clarendon Laboratory, Parks Road, Oxford OX1 3PU, United Kingdom}

\date{\today}



\begin{abstract}
We present a study on the optical losses of Fabry-P\'erot cavities subject to realistic transverse mirror misalignment. We consider mirrors of the two most prevalent surface forms: idealised spherical depressions, and Gaussian profiles generated by laser ablation. We first describe the mode mixing phenomena seen in the spherical mirror case and compare to the frequently-used clipping model, observing close agreement in the predicted diffraction loss, but with the addition of protective mode mixing at transverse degeneracies. We then discuss the Gaussian mirror case, detailing how the varying surface curvature across the mirror leads to complex variations in round trip loss and mode profile. In light of the severe mode distortion and strongly elevated loss predicted for many cavity lengths and transverse alignments when using Gaussian mirrors, we suggest that the consequences of mirror surface profile are carefully considered when designing cavity experiments. 
\end{abstract}

\maketitle

\section{Introduction}
\label{sec: introduction}

Fabry-P\'erot optical cavities are a leading platform for enhancing and controlling light-matter interactions, enabling coherent interactions between quantised emitters and single photons~\cite{Kimble:08}. This capability has been used to demonstrate deterministic single-photon production~\cite{Keller:04}, atom-photon logic gates~\cite{Reiserer:14} and remote entanglement generation~\cite{Stute:12}. The properties of the cavity constitute the core limitation to the success and scalability of many of their applications \cite{Goto:19, Schupp:21, Gao:23} and, accordingly, cavity design and fabrication remains an active area of research~\cite{Takahashi:14, Brekenfeld:20, Pscherer:21}
 
Many seminal demonstrations in optical cavity QED used cavities with superpolished mirrors~\cite{Thompson:92, Mabuchi:96, Hood:98, Ye:99}. These mirrors are typically highly spherical within the milling diameter~\cite{Rempe:92} and continue to be used for leading experiments~\cite{Schupp:21, Krutyanskiy:23}. However, to improve the strength and efficiency of light-matter interfaces, cavities with highly curved mirrors and thus low optical mode volume are beneficial \cite{Durak:14}, and alternative mirror fabrication techniques have been developed to produce the tight curvatures required while maintaining high surface quality~\cite{Trupke:05, Steinmetz:06, Dolan:10}.

One commonly-employed method is laser ablation, during which evaporation and surface tension effects produce highly-curved substrates with low roughness~\cite{Hunger:10, Rochau:21, Doherty:23}. This technique is often used to place micromirrors on the tips of optical fibres~\cite{Hunger:12}. Cavities constructed in this manner have been coupled to many promising emitters for quantum information processing, including neutral atoms~\cite{Barontini:15, Brekenfeld:20, Macha:20}, ions~\cite{Takahashi:20, Kobel:21}, quantum dots~\cite{Miguel-Sanchez:13} and nitrogen vacancies~\cite{Albrecht:14, Kaupp:16, Reidel:17}. However, the ablation laser generally imparts its Gaussian transverse intensity distribution into the mirror profile~\cite{Muller:10}, manifesting particular consequences for the emitter-photon system that are not observed with spherical or parabolic mirrors. Firstly, ellipticity of the addressing laser leads to anisotropic profiles and thus geometric birefringence~\cite{Uphoff:15}, which can introduce intracavity polarisation rotation~\cite{Barrett:19} that may frustrate applications such as remote entanglement generation~\cite{Kassa:22}. Secondly, the non-spherical mirror surface causes the well-known modes of a cavity with spherical mirrors~\cite{Siegman:86} to mix with each other upon reflection \cite{Kleckner:10}, forming new cavity eigenmodes with distinct transverse profiles that produce significant optical losses for certain geometries~\cite{Benedikter:15, Ott:16, Ruelle:22}. The issues related to the Gaussian shape of ablated mirrors have encouraged the development of more advanced ablation techniques~\cite{Takahashi:14, Ott:16} or the use of alternative fabrication methods~\cite{Hessenauer:23}.

Mode hybridisation in optical cavities has been well-studied for the idealised scenario of perfectly transversely aligned mirrors~\cite{Podoliak:17}. However, the mirrors of an optical cavity are commonly subject to transverse misalignment, whether induced by the mirror milling process, manufacturing tolerances of the mirror substrates, or the alignment and fixing of the mirror substrates relative to each other~\cite{Brandstatter:13, Saavedra:21}. It is therefore important to understand the combined impact of mirror profile and transverse misalignment to design optical cavity systems that can function reliably under realistic misalignment.

Here, we use recently developed extensions~\cite{Hughes:23_1} to the mode mixing method of Kleckner et al.~\cite{Kleckner:10} to model transverse misalignment with reduced numerical difficulty compared to conventional techniques. The paper is organised as follows. In Sec.~\ref{sec: theory summary}, we summarise the theory utilised in our investigation. In Sec.~\ref{sec:spherical}, we present results for cavities with finite-diameter spherical mirrors, comparing the calculated losses to the classical clipping model. We then analyse cavities with Gaussian-shaped mirrors in Sec.~\ref{subsec: Gaussian mirror results}, exploring how the variable surface curvature yields more complicated manifestations of mode-mixing physics. Finally, in Sec.~\ref{sec: Conclusion}, we suggest the implications of the results presented on the design of Fabry P\'erot microcavities.  

\section{Theory summary}
\label{sec: theory summary}
This paper compares the behaviour of both finite-diameter spherical mirrors (henceforth `spherical cap mirrors'), and Gaussian-shaped mirrors, under transverse misalignment. The analysis performed uses first a simple geometric approach for predicting the propagation direction and central waists of the resonant cavity modes under mirror misalignment, and then uses the mode mixing method \cite{Kleckner:10} to calculate the resonant modes more accurately. A summary of the geometric and mode mixing approaches to determining the cavity modes are given below.

\subsection{The geometric picture}

The geometric picture estimates the cavity mode in a simplified manner by restricting itself only to fundamental Gaussian beams. Firstly, the mode axis is chosen to be the line that intersects both mirrors orthogonal to their surface. Secondly, the transverse structure of a fundamental mode is determined by requiring that the wavefront curvature of the predicted mode matches the local curvature of the mirror at the intersection of mode and mirror for both mirrors~\cite{Blows:98}. This procedure determines the positions and the waists of the mode in each transverse direction.

The mode predicted by this method, henceforth known as `the geometric prediction' and denoted $\psig$, is useful to understand the impact of mirror geometry on mode propagation direction and central waist. However, this method accounts for the mirror shape only through its local gradient and curvature; higher order components in the Taylor expansion of the surface profile about the intersection point, which become important to describe the surface profile away from the central intersection with the mode, are not accounted for. For spherical cap mirrors, the mirror curvature remains constant within the mirror diameter, and therefore, provided the mode axis intersects the mirror within the finite diameter, the local curvature at intersection remains constant and only the propagation direction changes upon misalignment~\cite{Yariv:91}. However, for Gaussian-shaped mirrors, the local curvature varies across the surface, with a reduced and elliptical curvature away from the centre. Thus for cavities with misaligned Gaussian mirrors, the local curvature takes two principal values, both smaller than the central curvature of the mirror, but most strongly reduced in the direction of misalignment. Detailed algebraic and numeric results for the case of Gaussian-shaped mirrors may be found in Hughes et al.~\cite{Hughes:23_1}.

\subsection{Mode mixing method}

To understand the impact of the full shape of the mirror, a more complete method that can account for the entire surface profile is required. A variety of methods have been developed for this purpose, which are also relevant for optical interferometers~\cite{Bond:16}. These methods include the iterative diffraction integral technique to determine lowest loss~\cite{Fox:63} and higher order modes~\cite{Fox:68}, or more recently the discrete linear canonical transform to calculate the effect of a cavity round trip in the position basis~\cite{Ciobanu:21}. This investigation employs the mode mixing method~\cite{Kleckner:10}, which has been used to analyse the outcomes of microcavity experiments~\cite{Benedikter:15, Walker:21}. At a general level, this method describes the action of a mirror through the scattering of input modes to output modes in a Hermite-Gauss or Laguerre-Gauss basis, encoding this information as a matrix. A brief overview of the principle of the method and the basis functions will be given below.

In a Fabry P\'erot cavity, Maxwell's equations are typically simplified by assuming that the propagating field is beam-like and directed at small angles to the nominal $z$ axis. After this assumption, known as the paraxial approximation, the (assumed monochromatic) electric field can be described through a simpler scalar field $u^{\pm}(x,y,z)$ \cite{Barre:17} satisfying
\begin{equation}
    \bm{E}(x,y,z,t) = \bm{\epsilon} u^{\pm}(x,y,z)\exp(\mp ikz)\exp(i\omega t),
\label{eq: vector mode to scalar}
\end{equation}
where $\omega$ is the angular frequency, $k=\omega/c$ the wavevector, $\bm{\epsilon}$ the constant linear polarisation of the field, which must lie perpendicular to the $z$-axis, and $\pm$ denotes propagation towards positive or negative $z$ respectively. The function $u^{\pm}(x,y,z)$ must satisfy the paraxial wave equation
\begin{equation}
    \pdv{z}u^{\pm}(x,y,z) = \mp\frac{i}{2k}\left(\pdv[2]{x} + \pdv[2]{y}\right)u^{\pm}(x,y,z).
    \label{eq: paraxial equation}
\end{equation}

A particular set of solutions to this equation, which suit the boundary conditions imposed by spherical cavity mirrors, are the Hermite-Gauss solutions, which for symmetric cavities narrow to a central waist of width $w_0$ in the plane $z=0$. The individual solutions are indexed by the $x$ and $y$ indices $n_x, n_y \in \mathbb{N}$ respectively and are written
\begin{equation}
\begin{aligned}
u^{(\pm)}_{n_x,n_y}\left(x,y,z\right)  =  & a(z) H_{n_x}\left(\frac{\sqrt{2}x}{w(z)}\right)H_{n_y}  \left(\frac{\sqrt{2}y}{w(z)}\right) \\
\exp\left[-\frac{x^2+y^2}{w(z)^2}\right] & \exp\left[\mp ik\frac{x^2+y^2}{2R_u(z)}\right]\exp\left[\pm i(n_x+n_y+1)\Phi_G\right], \\
\end{aligned}
\label{eq: Hermite Gauss mode}
\end{equation}
where
\begin{equation}
    \begin{aligned}
        a(z)  =  \frac{1}{w(z)} \sqrt{\frac{2}{\pi}\frac{1}{2^{n_x+n_y} n_x!n_y!}}, & \quad
w(z)  = w_0 \sqrt{1+\left(\frac{z}{z_0}\right)^2}, \\
z_0 = \frac{\pi w_0^2}{\lambda}, \quad
R_u(z)  & = z\left(1+\left(\frac{z_0}{z}\right)^2\right), \quad
\Phi_G(z)  =  \arctan\left(\frac{z}{z_0}\right),
    \end{aligned}
\end{equation}
where the wavelength $\lambda=2\pi/k$, $H_i$ are the Hermite polynomials, and $z_0$ is the Rayleigh range of the beam. The set of solutions containing all $n_x$ and $n_y$ is complete and orthonormal for each transverse plane separately.

Mirrors normal to the $z$-axis are described through matrices whose elements are scattering amplitudes from ingoing modes propagating in one direction to outgoing modes propagating in the reverse direction. These matrix elements are conventionally calculated through numerical integration, but in this investigation we use a faster operator approach~\cite{Hughes:23_1}.

Once the mirror matrices (denoted $A$ and $B$ for the two mirrors respectively) are calculated, the round trip matrix
\begin{equation}
    M = BAe^{-2ikL}
\end{equation}
may be found from the sequential action of both mirrors and the accumulated round trip phase. The eigenmodes $\ket{\Psi_i}$ of the round trip matrix are modes of the cavity. These modes will generally not be basis states if the mirrors $A$ and $B$ themselves scatter amplitude between basis states.

As discussed earlier in this section, the mode mixing method uses basis modes derived under the paraxial approximation, and the propagation of these modes should be modified for high divergence angles \cite{Lax:75, Yu:84}. In this manuscript, the most divergent mode presented has a divergence half angle of 7$^{\circ}$, and most are considerably below this level, so the overall transverse mode structure of most modes should be largely accurate for the majority of results, but will not be perfectly accurate. In addition, cavities with highly-curved mirrors and sufficient finesse may still exhibit resolvable mode splittings from non-paraxial effects even when the mode is not highly divergent \cite{Zeppenfeld:10, vanExeter:22, Koks:22_2}

In order to parameterise the mirror shapes for comparison, Gaussian-shaped mirrors have a depth profile
\begin{equation}
    f_G(x,y) = D\left\{1-\exp(-\frac{x^2+y^2}{w_e^2})\right\},
    \label{eq: Gaussian mirror profile}
\end{equation}
taken, by convention, to have a value of zero at the centre of the depression and take more positive values towards the edges of the mirrors, where $x$, $y$ are transverse coordinates on the mirror surface, $D$ is the depth of the depression, and $w_e$ is the $1/e$-waist of the Gaussian profile. The central radius of curvature of such a mirror is $R_c = w_e^2/2D$. The mode mixing matrices of the Gaussian-shaped mirrors can be calculated through numerical integration, but for this investigation are calculated through the techniques detailed in~\cite{Hughes:23_1}. 

The spherical cap mirrors are assumed to have constant curvature inside of their nominal diameter $D_M$, and be completely non-reflective outside of that diameter. The spherical cap mirror matrices are calculated by numerical integration of the overlap elements over the reflective region.

\subsection{Overview of data presented}

The studies presented in this paper took two identical concave mirrors, either spherical cap or Gaussian-shaped, of a specified central radius of curvature $R_c$ and calculated the cavity mode as a function of length $L$, defined as the axial length between the centre of the mirror depressions, and the mirror misalignment. For each cavity length, two mirrors were formed on-axis, and misalignment was then included by displacing them successively in opposite directions along the $x$-axis; for this investigation, the translation operator derived in~\cite{Hughes:23_1} was used. For each cavity length, the basis of calculation was chosen to match the theoretical modes for a cavity with perfectly-aligned spherical mirrors of the same central curvature as the trial mirrors. For each mirror shape, cavity length and misalignment, the mode mixing method produces a set of cavity eigenmodes $\set{\ket{\Psi_i}}$ with associated round-trip eigenvalues $\set{\gamma_i}$ that determine the round-trip losses $\loss{\mathrm{RT_i}}=1-\abs{\gamma_i}^2$ of the eigenmodes within the cavity. From these eigenmodes, a mode of interest must be selected. We choose the mode of interest to be the one that has the greatest overlap with the geometrical prediction $\ket{\Psi_{0,0}^G}$, as the geometrically expected mode possesses the same simple transverse structure as the fundamental mode of an ideal cavity, which bestows advantages in many applications. The choice of this approach is justified further in App.~\ref{app: choosing_mode_of_interest}. In our investigation, the overlap with the geometrically expected mode is calculated through the matrix rotation methods of~\cite{Hughes:23_1}, but could also be determined by numerical integration of the cavity mode function overlap.

This method of determining the eigenmodes has the additional benefit that it means certain basis modes need not be considered by symmetry. As mirror misalignment defines the $x$-direction, the cavity system remains mirror symmetric in the $y$-direction, and therefore cavity eigenmodes must have odd or even $y$-parity. The geometrical prediction $\ket{\Psi_{0,0}^G}$ has even $y$-parity, and thus the mode of interest may only be composed of even $y$-parity basis states. This symmetry is exploited here to reduce the number of matrix elements that must be calculated.

In addition to the magnitude, each complex eigenvalue $\gamma_i$ has a phase, which must be zero (modulo $2\pi$) for the eigenmode to be resonant. In a spectroscopy experiment, the probe frequency would typically be tuned to hit resonance, at which point the mode profile and loss could be examined. However, to reduce the computational time and difficulty of interpretation, all cavities in this investigation were studied at a single wavelength (\SI{1033}{\nano\metre}) under the assumption that the mode structure would deform negligibly were the probe frequency tuned to hit resonance. This is reasonable for our data (see App.~\ref{app: Wavelength dependence}), but for shorter cavities this may be less valid due to the increased frequency tuning required to cover one free spectral range. 

In order to separate the impacts of mode pointing and local curvature variation on the calculated mode, after calculation the mode coefficients were expressed in a basis with the same waist size and position as the calculation basis, but with direction of propagation matching the calculated eigenmode. If the chosen eigenmode propagates at angle $\phi_x$ from the $z$ axis towards the $x$ axis, the mode propagates along the unit vector $(\sin(\phi_x), 0, \cos(\phi_x))$, and the "co-propagating basis" is such that the state $\ket{\Psi_{n_x, n_y}^C}$ has a corresponding cavity mode function $u^{(\pm)}_{n_x,n_y}\left(x_m,y_m,z_m\right)$, where
\begin{equation}
    x_m = x \cos(\phi_x) - z\sin(\phi_x), \quad y_m = y, \quad z_m = z \cos(\phi_x) + x\sin(\phi_x)
\end{equation}
are the mode coordinates, which are rotated from the standard Cartesian coordinates so that the mode propagates along the $z_m$ axis. Components of the cavity eigenmode can be expressed in the co-propagating basis using the rotation matrix methods presented in~\cite{Hughes:23_1}, but could equally be found through numerical integration of the overlap between the cavity eigenmode and co-propagating basis state. Note that this procedure does not present the cavity eigenmode in the basis of the geometrically expected mode, but instead the cavity eigenmode in a basis with the same waist as the calculation basis, rotated to match the eigenmode found. 

Cavities with spherical cap mirrors were simulated on a basis of the first 50 modes in the $x$-direction, and, using the $y$ mirror symmetry discussed above, the first 25 even modes in the $y$-direction, making a total of 1250 modes. For cavities with Gaussian shaped mirrors, the basis contained the first 100 modes in the $x$ direction, and the first 50 even modes in the $y$ direction, making a total of 5000 modes. In order to correctly model diffraction losses in the numerical method utilised, the calculation of the Gaussian profile initially occurred in a larger basis, containing 115 and 65 states in the $x$ and $y$ directions respectively, before being truncated to the calculation size as discussed in~\cite{Hughes:23_1}. Note that numerical integration techniques do not need this truncation step. These basis sizes were verified to produce convergence, and using larger bases yielded no significant changes to the results. For the purposes of comparing spherical cap and Gaussian-shaped mirror profiles, we use $2w_e$ of the Gaussian mirror as an analogue for the finite diameter $D_M$ of the spherical cap. This has the convenient implication that for a given central radius of curvature $R_c$ and diameter (either $D_M$ or $2w_e$), the spherical cap and Gaussian profiles have the same depth. Further details about the algorithm and its implementation are given in App.~\ref{app: algorithm}.

Finally, it should be noted here that, in this investigation, we study concave-concave cavities, which have been employed in many experiments \cite{Stute:12, Takahashi:17}, and are particularly useful when coupling the cavity field to an emitter which must remain distant from the mirror surfaces. However, the results presented can also be applied to plano-concave cavities, which have the advantage that the mirrors cannot be transversely misaligned from each other, and find application in a variety of contexts \cite{Wang:17, Malmir:22}. By symmetry arguments, the round trip loss of the plano-concave cavity with a lossless, infinite planar mirror is half the round trip loss of the concave-concave cavity with twice the length for zero transverse misalignment. No further consideration will be given to plano-concave designs for the remainder of the manuscript.


\section{Spherical Mirror Cavities}
\label{sec:spherical}

In the absence of finite diameter effects, cavities with spherical mirrors constitute an ideal case in which the behaviour of the cavity mode under transverse misalignment is treated in standard theory \cite{Boyd:61, Siegman:86}. Here, the mode angle tilts as the mirrors are misaligned, but the mode retains its Gaussian transverse intensity profile due to the uniform mirror curvature. It is expected that finite diameter spherical mirrors will follow this behaviour until the mirror misalignment is sufficient for a significant proportion of this predicted mode to fall outside of the mirror diameter.

\subsection{Loss structure}
\begin{figure}
\captionsetup{width=0.95\textwidth}
\centering\includegraphics[width=0.9\columnwidth]{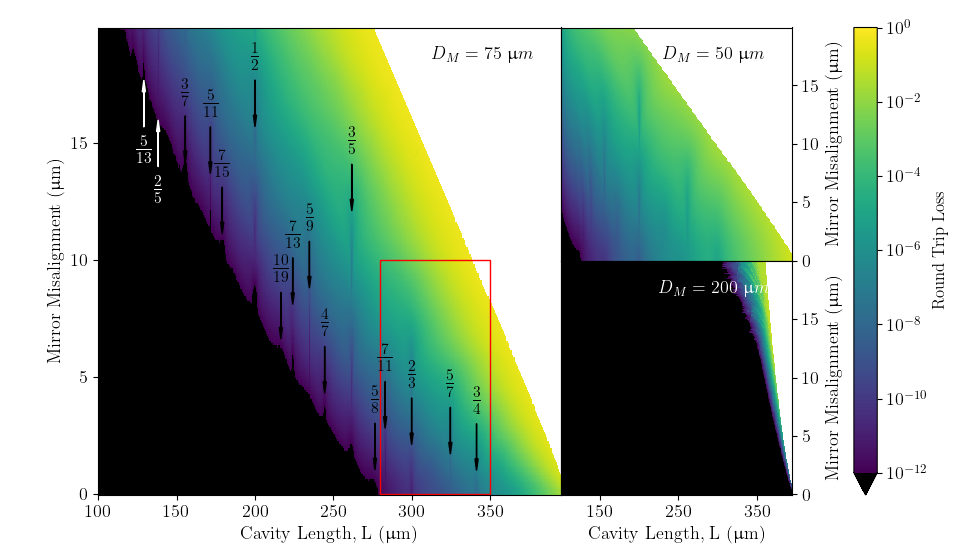}
\caption{Round trip loss for cavities with spherical cap mirrors as a function of cavity length and mirror misalignment. Each pair of mirrors forming the cavities has a central radius of curvature of $200~\upmu\mathrm{m}$, with data shown for three different diameters $D_M$ of the mirrors. The cavities are analysed with a wavelength of 1033~nm. Data is not calculated for translation-length combinations where the region inside one waist of the expected mode $\psig$ would not be fully enclosed within the spherical cap, with this region left white. Losses below $10^{-12}$ are not shown, as below this level numerical noise begins to become significant. In the larger plot, the sharp, low-loss bands are labelled by the integer ratio $q/p$ of transverse mode splitting to free spectral range to which the length corresponds. The red rectangle indicates the region taken for further analysis in figure~\ref{fig: spherical_panel}.}
\label{fig: spherical_shell_loss}
\end{figure}

To investigate the behaviour of cavities with spherical cap mirrors under transverse misalignment, cavity mirrors of radius of curvature $200~\upmu\mathrm{m}$ were modelled at three diameters and the cavity eigenmodes were calculated as a function of cavity length and transverse misalignment of the mirrors. The loss structure is presented in Fig.~\ref{fig: spherical_shell_loss}, qualitatively agreeing with the suggestion that the calculated diffraction loss arises when the mode encounters the finite diameter of the mirror.

One feature not predicted by the geometric model are the isolated bands of low loss for specific length values. These bands result from the resonant mixing of higher-order transverse modes with the geometrically-predicted fundamental Gaussian mode and can thus be associated with lengths at which particular transverse modes are degenerate. These degeneracies occur at lengths
\begin{equation}
    L_{p,q} = 2R\left[\frac{\tan^2\left(\frac{\pi q}{2 p}\right)}{1+\tan^2\left(\frac{\pi q}{2 p}\right)}\right]
\end{equation}
for integer $p$ and $q$, with $q/p$ the ratio of the transverse mode splitting to the free spectral range (in a cavity with zero diffraction loss and the same mirror curvature)~\cite{Arnaud:69}. It should be noted though that while these resonances reduce the round trip loss, for cases where the geometrically predicted mode remains largely inside the finite diameter, the impact on the mode shape is generally minimal.

\begin{figure}
\captionsetup{width=0.95\textwidth}
\centering\includegraphics[width=0.77\columnwidth]{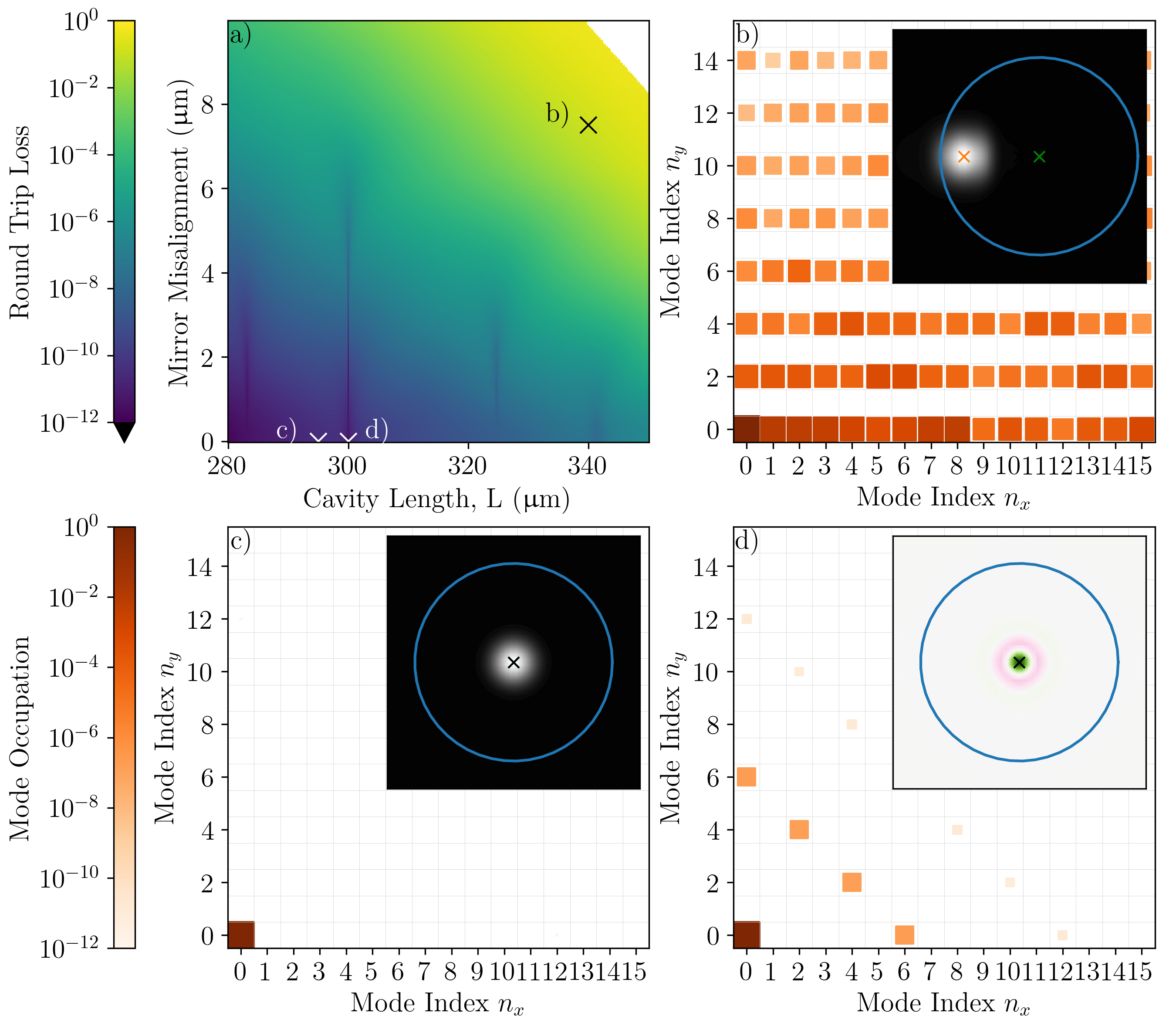}
\caption{Example round trip loss and cavity eigenmode data from cavities with spherical cap mirrors of diameter $D_M=75~\upmu\mathrm{m}$. a): The round trip loss for different cavity configurations, marking on 3 configurations of interest explored further in the corresponding panels. b) c) and d): breakdowns of the occupations of the cavity mode in the co-propagating basis $\copropagatingbasis$ at the configurations of interest, with insets depicting the mode in the plane of the mirror at positive $z$. For b) and c), the mode intensity is plotted, and for d) the difference in intensity compared to the geometric prediction $\psig$ is shown. The circle imposed on these insets depicts the mirror boundary. b): The high-loss mode formed as the light begins to impinge on the flats of the mirror. c): The mode in a non-misaligned case away from sharp dips in losses d): The mode in a non-misaligned case at a resonant reduction in losses. The intensity residuals indicate the mode on the mirror is more compact than the geometrical expectation $\psig$.}
\label{fig: spherical_panel}
\end{figure}

The onset of clipping loss and the role of transverse resonances in reducing these losses are investigated further in Fig.~\ref{fig: spherical_panel}. Firstly, the cavity becomes higher loss as the mode approaches the boundary of the spherical mirror. In the mode-mixing description, this loss manifests as a cascade of occupation to ever higher order modes (Fig.~\ref{fig: spherical_panel}b). Secondly, at the low loss bands, for example in Fig.~\ref{fig: spherical_panel}d), the higher-order transverse modes hybridise with the fundamental Gaussian mode, while there is very little hybridisation away from these resonances (Fig.~\ref{fig: spherical_panel}c). The modes of $\set{\ket{\Psi_{n_x,n_y}^C}}$ that hybridise can be predicted from the resonance label $q/p$ and from symmetry considerations. First, the resonance label (in the case studied $q/p=2/3$) determines the higher order modes that are resonant with the expected mode. These are the modes for which excitation index $\mathcal{I}=n_x+n_y$ is a multiple of $p$. Secondly, symmetry constrains that, at zero misalignment (which is the case presented), only modes with even $n_x$ and $n_y$ indices have both the $x$ and $y$ parity required to overlap with $\ket{\Psi_{0, 0}^{C}}$. Therefore, at zero misalignment and at the $q/p=2/3$ resonant length, $\ket{\Psi_{0, 0}^{C}}$ mixes with higher order modes for which $\mathcal{I}=n_x+n_y$ is a multiple of 6 and both $n_x$ and $n_y$ are even, as seen in the mode occupation patterns (Fig.~\ref{fig: spherical_panel}d). The accompanying intensity residual plot confirms that the mode hybridises to become physically more compact on the mirror, providing a mechanism for the observed reduced clipping loss.

\subsection{Comparison with classical clipping approximation}

A frequently-used method \cite{van_Dam:18, Flagan:22, Karpov:22} of estimating the losses induced by finite mirror diameter is the clipping loss approximation~\cite{Hunger:10}. This method calculates the round-trip loss as the power falling outside of the bounds of the mirrors during one round trip, on the assumption that the mode shape is unaffected by the power loss \cite{Clarke:18}. Extending the treatment of \cite{Hunger:10} to the off-axis case as performed in \cite{Gao:23}, the clipping loss is calculated through
\begin{equation}
\loss{clip} = 1- \left(\int_{S_M}\abs{u^{(G)}(x,y,z)}^2 \, dA\right)^2,
\end{equation}
where $u^{(G)}(x,y,z)$ is the cavity mode amplitude predicted by the geometric model, and $S_M$ is the mirror surface such that $\int_{S_\infty}\abs{u}^2 \, dA = 1$, where $S_\infty$ is the surface of an infinite mirror. The squared integral in the expression for $\loss{clip}$ accounts for the two reflections per round trip.

\begin{figure}
\centering\includegraphics[width=0.85\columnwidth]{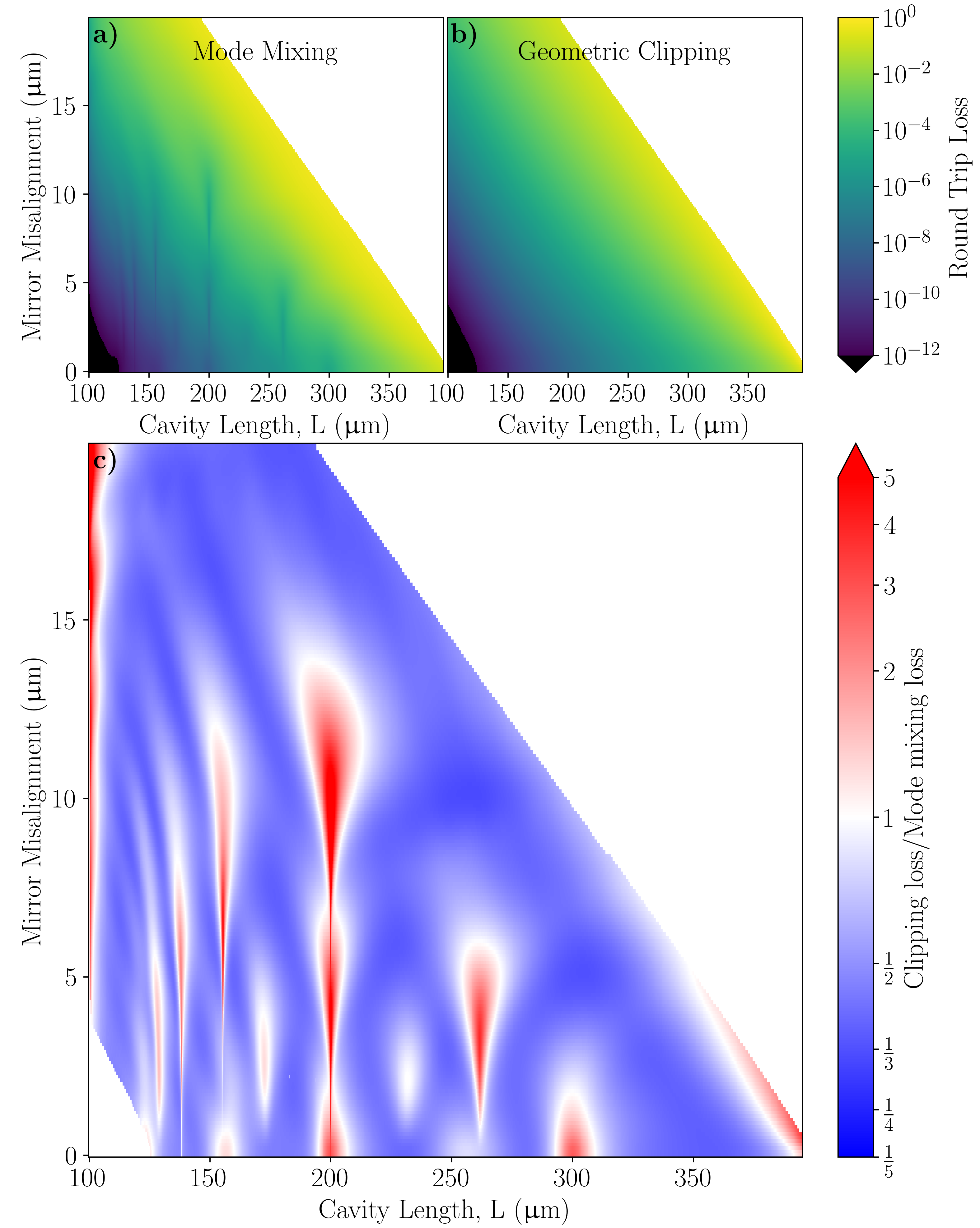}
\caption[Comparison of mode mixing and clipping calculations of round-trip loss]{Comparison of the round trip loss of cavities with spherical cap mirrors of diameter $D_M=50~\upmu$m predicted through both the mode mixing method and classical clipping approximation. a) The round-trip loss as a function of cavity length and misalignment for mode mixing calculations and b) classical clipping calculations. c) The ratio of the classical clipping loss to the mode mixing loss on a log scale. Red indicates that clipping loss exceeds that calculated by mode mixing, blue indicates the opposite, and white that the methods agree. Data is not shown for the case where the loss determined by either method is below $10^{-12}$, as these results are vulnerable to numerical noise.}
\label{fig: spherical clipping loss comparison}
\end{figure}

The round trip losses predicted by the clipping approximation and mode mixing method are compared in Fig.~\ref{fig: spherical clipping loss comparison}. Generally, the clipping loss approximation underestimates the cavity loss, although the scale of the underestimate remains within an order of magnitude throughout. The biggest disparities between the methods occur at the transverse resonances, where the clipping loss approximation overestimates the loss because the loss is reduced by transverse mode mixing, which the clipping approximation cannot invoke. For particular configurations the difference can surpass a factor of 10. Overall, the clipping loss approximation is sufficient to estimate the round trip loss within an order of magnitude, with the exception of configurations of significant mixing, for which the clipping loss estimate is conservative.

\section{Gaussian Mirror Cavities}
\label{subsec: Gaussian mirror results}
We now discuss mode hybridisation in cavities with misaligned Gaussian-shaped mirrors. Due to the differences between Gaussian and spherical cap mirror profiles, the concepts and terminology used to understand spherical cap mirrors in Sec.~\ref{sec:spherical} must be adapted. Firstly, while a spherical cap profile has a single fixed curvature within its finite diameter, a Gaussian-shaped mirror has a variable curvature across its surface, introducing a distinction between the central radius of curvature on the axis of the mirror, and the local radius of curvature where the mode intersects the mirror. The expected mode $\psig$ for the Gaussian case must account for the local curvature of the mirror, and therefore, at finite misalignment, the expected mode differs between spherical cap and Gaussian-shaped mirrors of the same central radius of curvature, though it remains a fundamental Gaussian beam. Secondly, while the spherical cap mirror profile becomes abruptly non-concave at the finite diameter, the concavity of the Gaussian profile gradually reduces away from the centre. Nevertheless, in the Gaussian-shaped case, there remains a boundary outside of which the mirror is not concave.

\begin{figure}
\captionsetup{width=0.95\textwidth}
\centering\includegraphics[width=0.9\columnwidth]{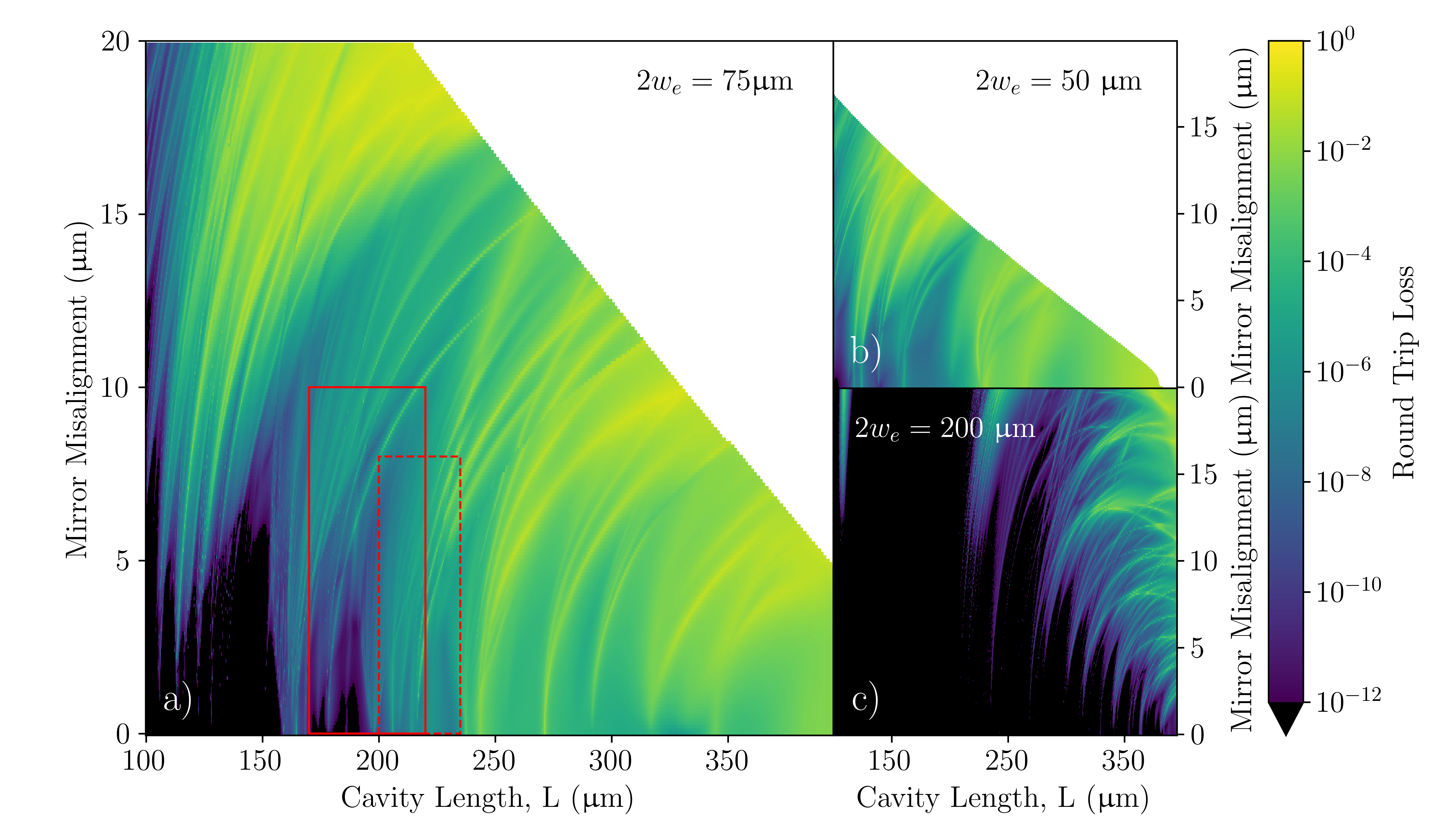}
\caption{Round trip loss for cavities with Gaussian mirrors as a function of cavity length and mirror misalignment. Each pair of mirrors forming the cavities has a central radius of curvature of $R_c=200~\upmu\mathrm{m}$, but three different Gaussian widths (as marked on the plots) and thus depths are shown. The cavities are analysed with a wavelength of \SI{1033}{\nano\metre}. Data is not shown for cases where the region inside one waist of the expected mode $\psig$ would not be fully enclosed within the positive curvature region of the Gaussian mirror, with this region left white. Losses below $10^{-12}$ are not shown, as below this level numerical noise begins to become significant. The solid (dashed) red boxes in a) indicate regions of interest that will be explored in figures \ref{fig: gaussian_7_resonance_example_panel} and \ref{fig: gaussian_4_resonance_example_panel}.}
\label{fig:gaussian_loss_map}
\end{figure}

The continuously-varying curvature of the Gaussian profile leads to more complicated structures in the dependence of round trip loss on cavity configuration, as exemplified in Fig.~\ref{fig:gaussian_loss_map}. The most striking visual element are bands of high loss, increasing in prevalence as the mirror diameter is reduced. 

\begin{figure}
\captionsetup{width=0.95\textwidth}
\centering\includegraphics[width=0.95\columnwidth]{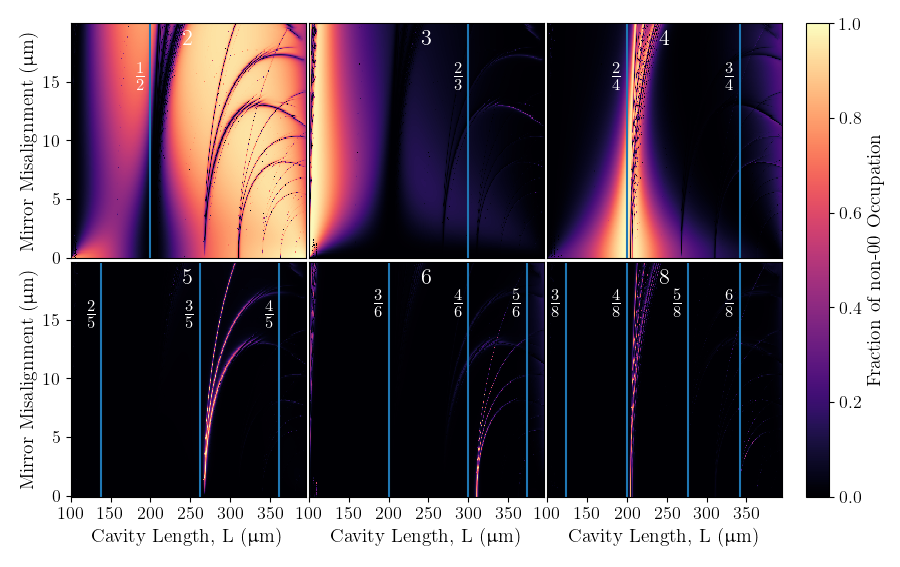}
\caption{Occupation of different excitation indices $\mathcal{I}$ of the cavity mode in the co-propagating basis $\left\{\ket{\Psi_{n_x,n_y}^{C}}\right\}$ over the length-misalignment map for Gaussian mirrors of $2w_e=200~\upmu\mathrm{m}$ and central radius of curvature $200~\upmu\mathrm{m}$. Each panel corresponds to the labelled excitation index $\mathcal{I}$, with the lengths of resonance of each excitation index with the $\ket{\Psi_{0,0}^{C}}$ marked by vertical lines with the rational ratio $q/p$ labelled. For each $\mathcal{I}$, the proportion of its occupation out of all not in the $\ket{\Psi_{0,0}^{C}}$ is plotted. As the mirror misalignment increases these resonances move to longer lengths and split into multiplets.}
\label{fig:mode_families}
\end{figure}

It is easiest to understand the physics behind these features in the $2w_e=200~\upmu\mathrm{m}$ case, depicted in Fig.~\ref{fig:mode_families}, where the mirror has a relatively large Gaussian width and thus deviates minimally from the spherical profile for a large region about its centre.  As observed in~\cite{Benedikter:15, Koks:22_1}, occupation of higher order transverse modes is associated with mode distortion and elevated loss, and typically occurs at degeneracies between the high order modes and the fundamental. In a perfect spherical cavity, mode degeneracy conditions are determined by the sum of transverse indices $n_x$ and $n_y$, and therefore we categorise the mode intensity in the co-propagating basis $\copropagatingbasis$ according to `transverse excitation' $\mathcal{I}=n_x+n_y$, finding that resonances are often dominated by a particular $\mathcal{I}$. The various behaviour seen in Fig.~\ref{fig:mode_families} can largely be understood through mode degeneracy and symmetry, as for the spherical cap case, with the more complex behaviour a consequence of the variable curvature across the Gaussian mirror. In the subsequent sections, the individual aspects of the loss structures are discussed in turn. 

\subsection{Mode degeneracy shifts}
\begin{figure}
\captionsetup{width=0.95\textwidth}
\centering\includegraphics[width=0.95\columnwidth]{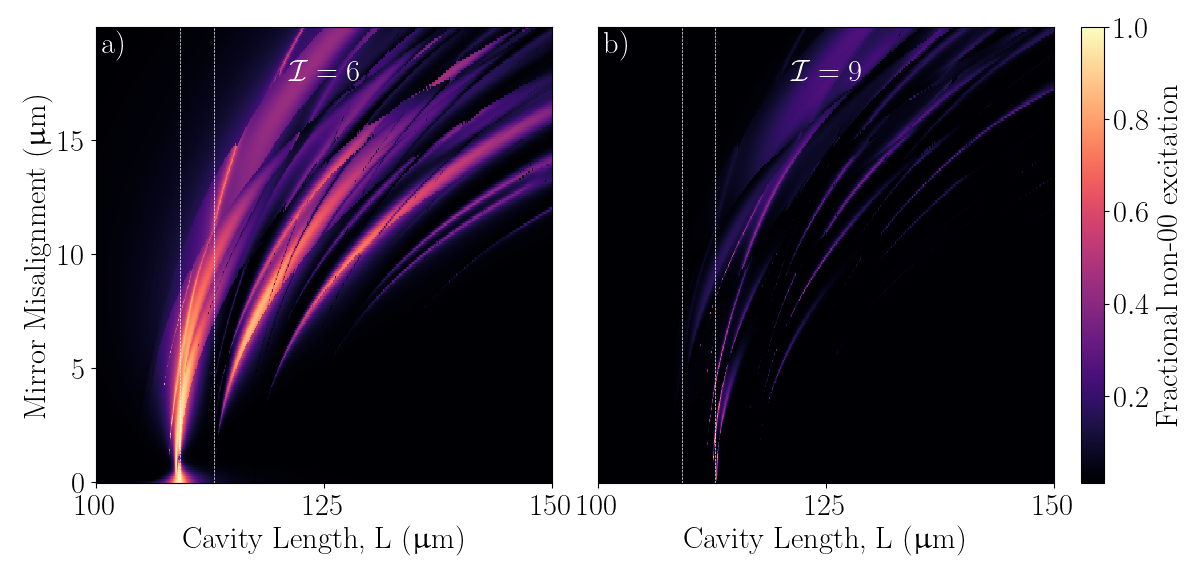}
\caption{Example of the resonance shift dependence on excitation index $\mathcal{I}$. Fraction of the non-00 occupation of $\copropagatingbasis$ in a) $\mathcal{I}=6$ and b) $\mathcal{I}=9$ as a function of cavity length and mirror misalignment for cavities with Gaussian-shaped mirrors of $1/e$-diameter $2w_e=$~\SI{75}{\micro\metre}. The strongest occupation of $\mathcal{I}=6$ tends to occur at lower lengths than for $\mathcal{I}=9$, as can be judged using the guide lines (white, dotted), which are in the same position on each plot}
\label{fig: across excitation index excitation panel}
\end{figure}

\begin{figure}
\captionsetup{width=0.95\textwidth}
\centering\includegraphics[width=0.95\columnwidth]{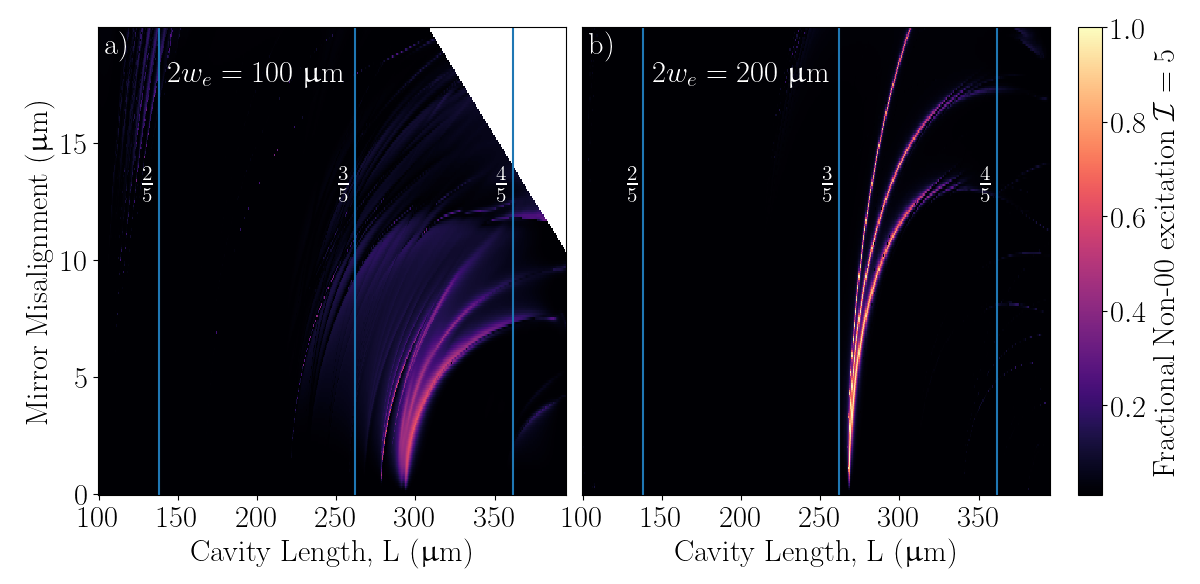}
\caption{Example of the resonance shift and its dependence on the Gaussian mirror width. The non-00 probability proportion in $\mathcal{I}=5$ of $\copropagatingbasis$ is shown for a cavity with Gaussian mirrors of central radius of curvature $R_c=$~\SI{200}{\micro\metre} and $1/e$-diameter $2w_e$ of a) \SI{100}{\micro\metre} and b) \SI{200}{\micro\metre}. Focusing particularly on the resonant mixing around $p/q=3/5$, it is seen that the shift of this resonance to longer lengths than for cavities with spherical mirrors is much more pronounced for the mirror with the smaller $w_e$ (panel a).}
\label{fig: across gaussian waist excitation panel}
\end{figure}

In analogy to the low loss bands observed with spherical mirrors, the high loss bands in Fig.~\ref{fig:gaussian_loss_map} can be attributed to degeneracy of the fundamental and higher-order transverse modes. For spherical cap mirrors, the cavity lengths at which mixing features occurred were precisely the lengths of transverse mode degeneracies in an ideal spherical mirror cavity. However, for Gaussian mirrors, loss bands are generally shifted to greater cavity length values than expected, both with and without mirror misalignment. This is due to the distributed intensity of the mode across the mirror, which means that the mode experiences an effective curvature that is some weighting of the local curvatures it encounters across the mirror. As the maximum local curvature is found at the centre of the mirror, the effective radius of curvature is always bigger than the nominal, central radius of curvature, and thus transverse resonances are shifted to longer lengths. As the mode order increases, a larger region of the mirror is explored by the mode, and the resonance length shift is greater, as seen in Fig.~\ref{fig: across excitation index excitation panel}. This contrasts with the degeneracy observed with spherical mirrors, where by example $\mathcal{I} = 3$, $\mathcal{I} = 6$ and  $\mathcal{I} = 9$ are coincident for $q/p$ = 1/3. Similarly, as the diameter of the Gaussian mirror is expanded, the cavity mode addresses a region that can be better approximated as spherical, shifting the loss bands back to their expected length value, as shown in Fig.~\ref{fig: across gaussian waist excitation panel}. 

\subsection{Ellipticity}

The resonances associated with high loss appear to both curve to higher cavity lengths and to split into multiplets as the mirrors are misaligned. These aspects can again be understood from changes in effective radius of curvature experienced by the mode upon reflection from the Gaussian mirrors. As summarised in Sec.~\ref{sec: theory summary} and discussed in more detail in \cite{Hughes:23_1}, the local curvature of the mirror at the intersection with the centre of the expected mode decreases as the mirror is misaligned, with the decrease much stronger in the direction of misalignment ($x$). The decrease in curvature pushes all transverse resonances to longer lengths as misalignment increases, rather than remaining at constant length as for the resonant features of the spherical cap mirror. At non-zero misalignment, the difference in radius of curvature in the $x$ and $y$ directions splits the resonant features into multiplets; within a given $\mathcal{I}$, the components with higher $n_x$ index are resonant at longer lengths. The multiplicity of the multiplets can thus be predicted. For example, for the $\mathcal{I}=6$ resonance, there are 7 states in the Hermite-Gauss basis, but the symmetry of the system about the $y$-axis dictates that the geometrically expected mode $\psig$ will only couple with the modes with even $n_y$-indices. Thus this resonance splits to a quadruplet, and the $\mathcal{I}=5$ peak should be a triplet, as seen in Fig~\ref{fig:mode_families}. Finally, it should be noted that the different effective curvatures in the $x$ and $y$ directions will also introduce a geometric birefringence \cite{Uphoff:15}, but this phenomenon is beyond the scope of the scalar mode mixing theory.

\begin{figure}
\captionsetup{width=0.95\textwidth}
\centering\includegraphics[width=0.95\columnwidth]{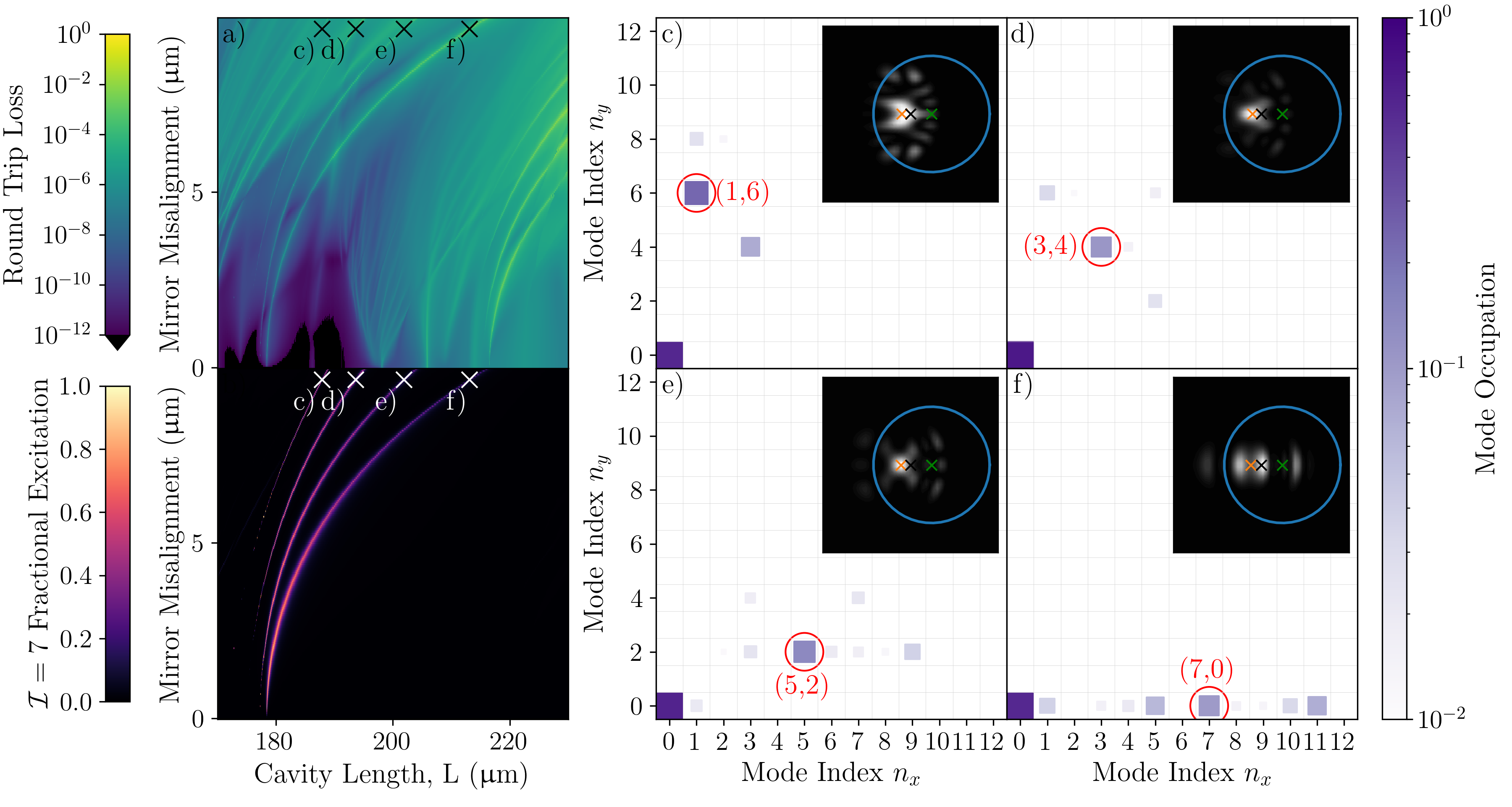}
\caption{Study of a resonance feature with modes of $\mathcal{I}=7$ for the $2w_e=75~\upmu\mathrm{m}$ mirrors. a) Cavity round-trip loss as a function of length and mirror misalignment and b) fraction of the non-$\ket{\Psi_{0,0}^{C}}$ occupation in $\mathcal{I}=7$, showing that the multiplet has strong occupation in this excitation class. On a) and b), four example points, corresponding to panels (c-f) respectively, are marked. (c-f): Mode compositions and profiles of the configurations shown in panels (a) (b). The main axis shows the occupation of the $\left\{\ket{\Psi_{n_x, n_y}^{C}}\right\}$ basis, with the inset showing the mode intensity on of one of the cavity mirrors. On each figure, the basis state with the largest non-$\ket{\Psi_{0,0}^{C}}$ occupation is ringed and labelled according to $(n_x, n_y)$, with this $n_x$ increasing across the multiplet in the direction of increasing length.}
\label{fig: gaussian_7_resonance_example_panel}
\end{figure}

To exemplify this physics, an $\mathcal{I}=7$ multiplet is studied in Fig.~\ref{fig: gaussian_7_resonance_example_panel}. As the misalignment increases, the feature splits into a quadruplet, determined by the number of even-$n_y$ states within $\mathcal{I}=7$. The mode composition of points in each of the four arms of the quadruplet was analysed. The $(n_x, n_y)$ indices of the dominant non-00 component in $\left\{\ket{\Psi_{n_x, n_y}^{C}}\right\}$ for each point was, in order of increasing resonant length at a given misalignment, ${(1,6)}$, ${(3,4)}$, ${(5,2)}$, and ${(7,0)}$. As expected, the resonances with higher $n_x$ occur at longer lengths, as these higher-order modes are most affected by the strong reduction in curvature in the direction of misalignment. Despite all features belonging to the $\mathcal{I}=7$ resonance, each peak presents a dramatically different mode shape, as the higher order mode that mixes most strongly changes between the peaks of the multiplet.

\subsection{Parity}
Some of the loss bands seen in Fig.~\ref{fig:gaussian_loss_map} emerge only at non-zero misalignment. As most easily seen in Fig.~\ref{fig:mode_families}, these bands correspond to odd values of $\mathcal{I}$, and the phenomenon is consequently understood through symmetry. Mode mixing occurs when the mirror mixes the $\ket{\Psi_{0,0}^{C}}$ mode (which has even-parity in both Cartesian directions) with a higher order mode. At zero misalignment, the Gaussian mirror has mirror symmetry in both transverse directions about the point where the mode intercepts the mirror, and therefore mixing can only occur with modes of even parity in both $x$ and $y$ directions. Such modes exist only for even $\mathcal{I}$. Therefore, features corresponding to mixing with odd $\mathcal{I}$ cannot extend to zero misalignment. At non-zero misalignment, the mode intersects the mirror away from its nominal centre. In the direction of the misalignment, the mirror is no longer symmetric about the intersection of the mode on the mirror due to the varying radius of curvature on either side of this intersection point.  This means that $\ket{\Psi_{0,0}^{C}}$ can couple into modes with both odd and even parity in the direction of misalignment, allowing for coupling into states with  odd $\mathcal{I}$. In this way, the resonant features can be classed according to odd or even symmetry in the direction of misalignment. The variable curvature of Gaussian mirrors renders the cavity mode vulnerable to odd-symmetry resonant features should the mirrors suffer residual transverse misalignment.

\subsection{Mode distortion}
\begin{figure}
\captionsetup{width=0.95\textwidth}
\centering\includegraphics[width=0.9\columnwidth]{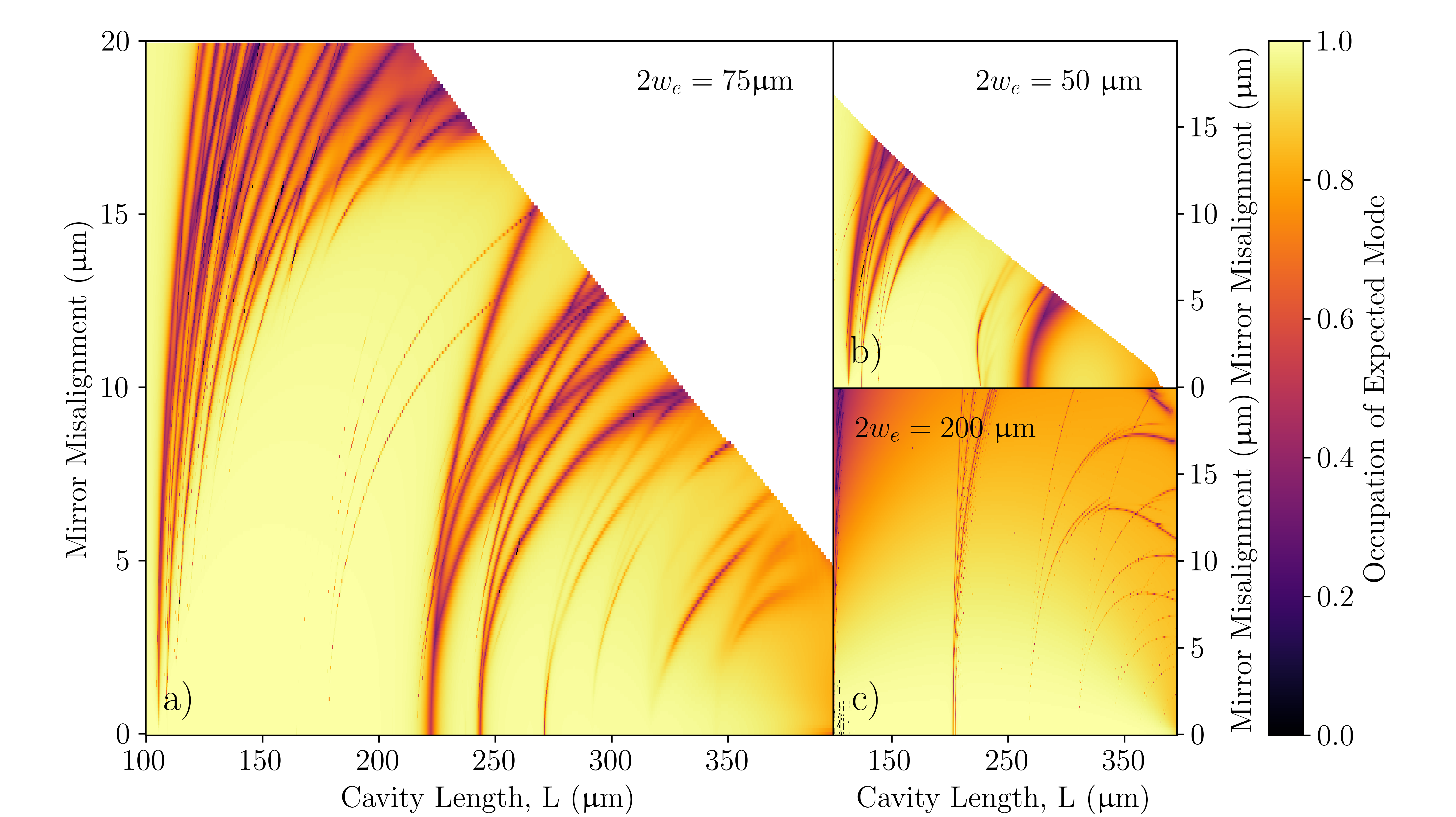}
\caption{Overlap of the selected cavity eigenmode with geometrically expected mode $\psig$ for cavities with Gaussian mirrors as a function of cavity length and mirror misalignment. Each pair of mirrors forming the cavities has a central radius of curvature of $R_c = 200~\upmu\mathrm{m}$, but the plots show three different Gaussian widths (as marked on the plots) and thus depths. The cavities are analysed with a wavelength of \SI{1033}{\nano\metre}. Data is not shown for cases where the region inside one waist of the expected mode $\psig$ would not be fully enclosed within the positive curvature region of the Gaussian mirror, with this region left white.}
\label{fig: gaussian_mirror_coupling}
\end{figure}

The mode of interest was selected by finding the cavity eigenmode with the greatest overlap with the geometrically expected mode $\ket{\Psi_{0,0}^{G}}$. This method was chosen on the assumption that there would usually exist a cavity eigenmode that was a perturbation of the geometrically expected mode, which retains the transverse structure of a fundamental Gaussian beam. However, as shown in Fig.~\ref{fig: gaussian_mirror_coupling}, while such a cavity eigenmode can typically be found, there are configurations where the chosen eigenmode has an overlap of approximately 50\%, or occasionally even less, with the expected mode, even for transversely aligned mirrors. This arises because the expected mode can fully hybridise with a higher order mode at transverse degeneracies, as observed in~\cite{Koks:22_1}, meaning that the cavity eigenmode is not a small perturbation of the expected mode.

\begin{figure}
\captionsetup{width=0.95\textwidth}
\centering\includegraphics[width=0.95\columnwidth]{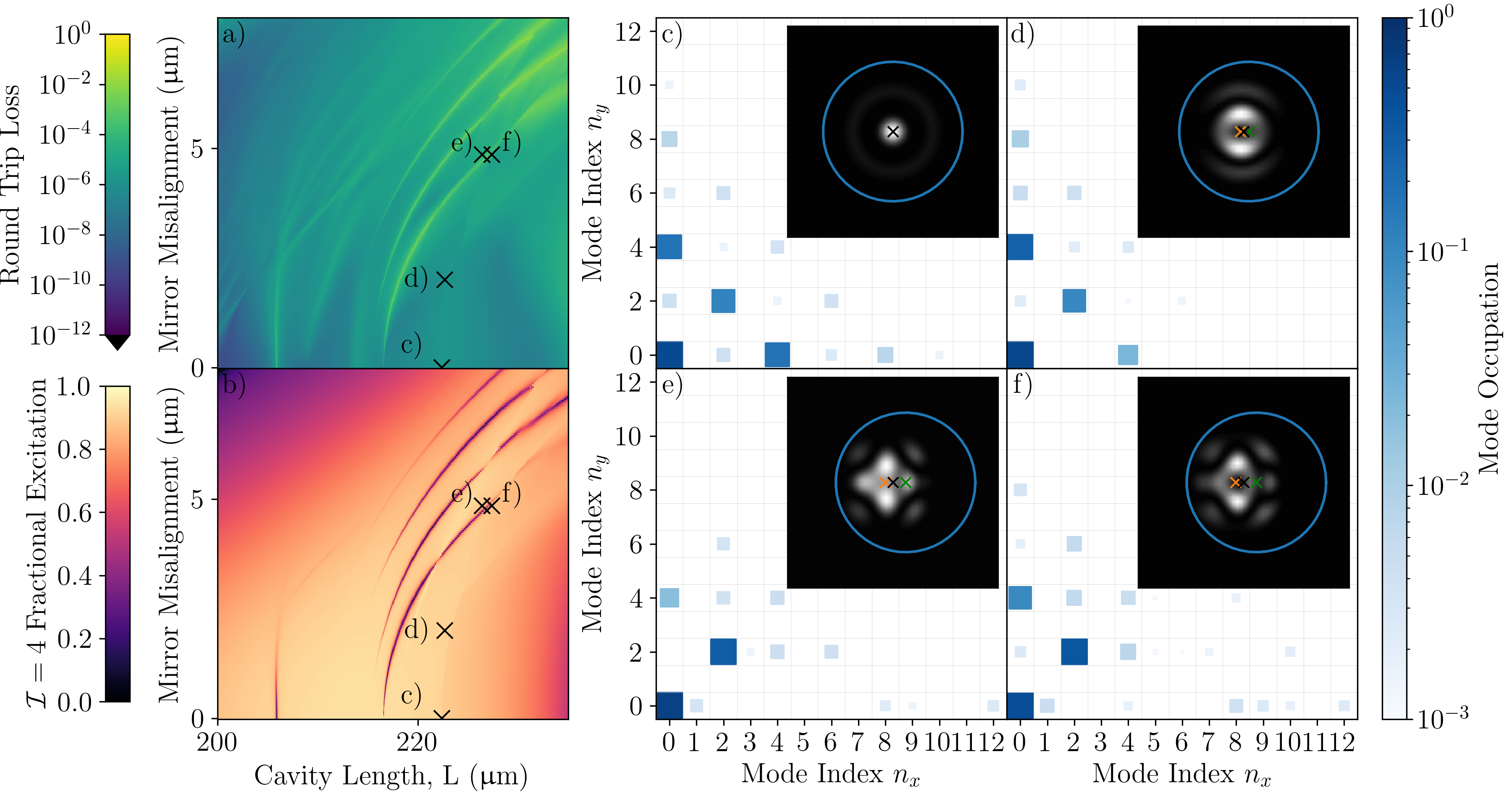}
\caption{Study of the confocal configurations of cavities with $2w_e=75~\upmu\mathrm{m}$ mirrors. a) Round-trip loss as a function of length and mirror misalignment. b) The fraction of the non-00 occupation of $\left\{\ket{\Psi_{n_x, n_y}^{C}}\right\}$ with $\mathcal{I}=4$; showing a very broad peak across the configurations shown.  On a) and b), four example points corresponding to (c-f) respectively, are marked. (c-f): The mode compositions on the mirror in $\left\{\ket{\Psi_{n_x, n_y}^{C}}\right\}$ for the configurations shown in panels (a) (b), with insets showing mode intensity.}
\label{fig: gaussian_4_resonance_example_panel}
\end{figure}

To investigate such cases further, in Fig.~\ref{fig: gaussian_4_resonance_example_panel}, we study the region around the confocal configurations of this system, where there is very strong occupation of $\mathcal{I}=4$ throughout, but with narrower, high loss regions within. Here, the occupation of higher order modes in the co-propagating basis is so strong as to make identifying the mode of interest challenging, as the geometrically expected mode hybridises very strongly, meaning no mode strongly resembles the expectation. The selected mode is still the one which maximises the overlap with the expectation, but, where the decision between eigenmodes is finely-balanced, the shape of the selected mode can change discontinuously across the length-misalignment space as different mode hybridisations are chosen. These highly distorted modes and discontinuous changes in the transverse profile can be seen around narrow, high loss features where the mode structure might be expected to be complex (Fig.~\ref{fig: gaussian_4_resonance_example_panel} e and f), but also for regions of relatively low loss (Fig.~\ref{fig: gaussian_4_resonance_example_panel} c and d). In applications requiring coupling the cavity mode to an emitter or the extraction of photons from the cavity to a single-mode fibre, strong mode distortion is, in itself, problematic. Therefore, for applications of cavities with Gaussian-shaped mirrors, it is important to consider the mode distortion as well as the round trip loss.

\subsection{Loss increase at mode degeneracy}
An obvious point of difference between the two mirrors shapes is that, with spherical cap mirrors, mixing at mode degeneracies leads to low loss features, whereas for cavities with Gaussian-shaped mirrors, these features have elevated loss. The spherical cap mirror surface can be partitioned into one section inside the diameter, which does not mix the co-propagating basis modes, and the region outside the mirror diameter, which causes mixing and loss. At mode degeneracies, the cavity eigenmode can hybridise to reduce the intensity falling on the lossy region, causing a reduction in round trip loss. For the Gaussian mirror case, the same argument cannot be used directly, because the mirror cannot be partitioned into mixing and non-mixing areas. While this is not a direct reason that the mixing-induced bands should be high loss for cavities with Gaussian-shaped mirrors, it does suggest that the mechanism by which loss was reduced for cavities with spherical-cap mirrors does not apply when those mirrors are Gaussian-shaped.


\section{Conclusion}
\label{sec: Conclusion}

We have conducted a numerical study into the round trip losses of cavities with spherical cap and Gaussian-shaped mirrors under transverse misalignment. The diffraction losses of cavities with spherical cap mirrors were found to broadly agree with the frequently-used classical clipping approximation. Deviations from the predictions of this approximation are seen as sharp dips in the round trip losses at lengths for which higher order transverse modes are degenerate with the expected mode. These higher order modes hybridise with the expected mode to reduce the intensity falling outside the finite diameter mirror and reduce the losses, but the overall deformation of the mode is not significant.

In the case of Gaussian mirrors, however, the variation in curvature across the mirror introduces complicated hybridisation effects that distort the mode and increase the cavity loss, even when the expected mode remains well inside the concave region of the mirror. Instead of low loss bands at particular resonant lengths, the resonant features are high loss bands which split and curve over to longer length as the cavity mirrors are misaligned. This behaviour can be understood through an `effective mirror curvature' seen by the cavity mode, which tends to be lower for higher order basis states, and to reduce away from the central axis (most prominently in the direction of translation, but also in the orthogonal direction). These two effects alter the positions of resonance, shifting them to higher cavity lengths for zero misalignment and curving to longer lengths with increasing misalignment, while splitting into multiplets. The multiplicity of the resonance can be predicted by counting the number of symmetry-allowed couplings of the relevant excitation index. These resonances possess either odd or even character, with the odd resonances only observable when the cavity is misaligned. For particular configurations, the mixing induced by the Gaussian mirror may be so strong that no cavity eigenmode resembles a fundamental Gaussian mode.

With regards to the use of Gaussian shaped mirrors for quantum technologies, our results indicate that care must be taken to ensure that the light matter interface would function as expected, which is not simply a case of ensuring that the expected mode lies within the confining part of the mirror. In contrast to the spherical cap case where the mixing features are beneficial, sparse, and do not distort the cavity mode significantly, the Gaussian mirror case has high-loss regions littered throughout the length-misalignment landscape. These regions can be broad (for example at lengths just exceeding the confocal length), or very sharp, and the cavity length values that result in high loss depend strongly on the shape of the mirror and the transverse misalignment. Misalignment, in addition to bringing the mode closer to the edge of the confining region of the mirror, brings an extra deleterious effect as the mode is also vulnerable to odd-character transverse resonances. We anticipate that these observations will find use in the selection of cavity construction techniques for future cavity QED experiments, and that the methods and techniques presented will advance understanding of losses in cavities with Gaussian-shaped mirrors. 
\section{Acknowledgements}
This work was funded by the UK Engineering and Physical Sciences Research Council Hub in Quantum Computing and Simulation (EP/T001062/1) and the European Union Quantum Technology Flagship Project AQTION (No. 820495). The authors would like to acknowledge the use of the University of Oxford Advanced Research Computing (ARC) facility in carrying out this work. http://dx.doi.org/10.5281/zenodo.22558. Data underlying the results presented in this paper are available in Ref.~\cite{Hughes:23_2}. The code that generated the data may be obtained from the authors at reasonable request.

\appendix

\section{Choice of Mode of Interest}
\label{app: choosing_mode_of_interest}
The mode mixing method used in the manuscript produces a set of cavity eigenvectors, and an associated set of eigenvalues which encode the losses. No mode has a privileged position within these sets, and therefore it is necessary to choose which mode is most interesting for a given application. This section discusses the merits of different methods to choose the mode of interest.

\begin{figure}[h]
\captionsetup{width=0.95\textwidth}
\centering\includegraphics[width=0.40\columnwidth]{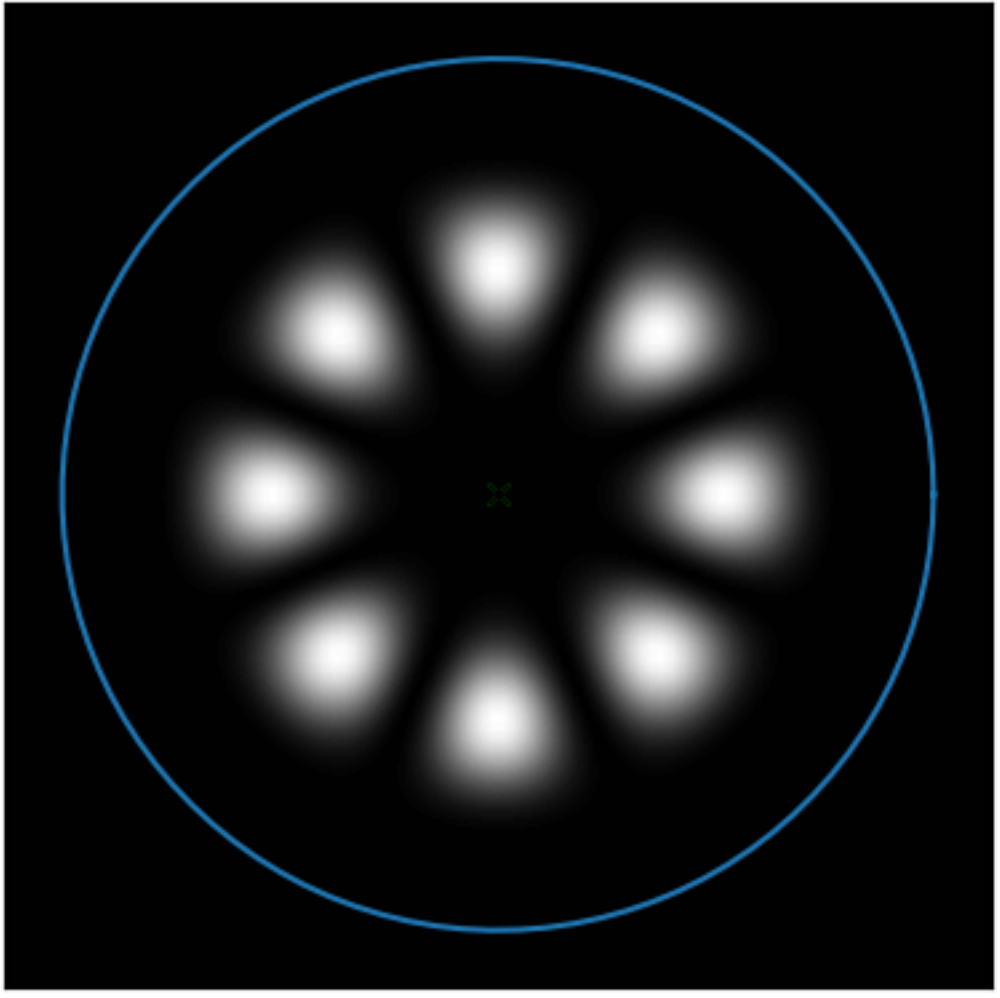}
\caption{The profile of the smallest loss cavity eigenmode on the mirror for the case of $2w_e=75~\upmu\mathrm{m}$ Gaussian mirrors of central curvature $R_c = 200~\upmu\mathrm{m}$ and cavity length of $L=238~\upmu\mathrm{m}$ with no transverse mirror misalignment interrogated 1033 nm. For this case, and for many other configurations explored, the lowest loss mode differs significantly in structure from a typical Gaussian beam.}
\label{fig: example_smallest_loss_not_fundamental}
\end{figure}

One could choose the mode with the lowest loss to be the mode of interest. This mode often strongly resembles the geometrically expected mode. However, when the cavity is misaligned strongly, or even for certain configurations in the absence of misalignment, the lowest loss mode can look very unlike the expected Gaussian beam (see Fig.~\ref{fig: example_smallest_loss_not_fundamental}). Such a mode would not be ideal for coupling to an emitter or easy to couple into a fibre. In this manuscript, we choose the mode of interest to be the one with the highest overlap with that expected from the geometrical approach. This encourages the selection of modes with simple transverse structure that are more amenable to most applications

A possible alternative is to choose the mode with the lowest $M^2$ value (through methods detailed in Nemes and Siegman~\cite{Nemes:94}). This method would select the mode with the smallest far-field diffraction, which should have the `simplest' spatial structure. The advantage of this method is that it is completely basis independent, and is associated only with the geometric properties of the eigenmode. In our investigation, we chose to use the overlap with the expected mode instead to bias the selection towards modes that propagate in the expected direction.

\section{Wavelength Dependence of Loss Features}
\label{app: Wavelength dependence}
The calculations in this manuscript were all conducted at a single wavelength (\SI{1033}{\nano\metre}). At this wavelength, the $j$th cavity eigenmode $\ket{\Psi_j}$ has an eigenvalue $\gamma_j = \abs{\gamma_j} e^{i\phi_j}$. The round-trip phase $\phi_j$ is then ignored, and the loss calculated as $(\mathcal{L}_{\text{RT}})_j= 1-\abs{\gamma_j}^2$. This generates a loss value consistent with the round-trip loss that would be seen if that wavelength were resonant. However, in a spectroscopy experiment, for a given cavity configuration the probe wavelength is scanned until the resonance is found. For this case it is unclear whether it is reasonable to calculate the results at a single wavelength in order to forego the extra computation entailed with calculating at several wavelengths. Changing the wavelength has two principal effects:
\begin{enumerate}
    \item Altering the round-trip phase $e^{-2ikL}$
    \item Changing the mirror matrix components
\end{enumerate}
The first effect is typically dominant, and is the simplest to account for. The phase factor added to the round-trip matrix will not alter the eigenvectors $\ket{\Psi_j}$, only the phases $\phi_j$ of the eigenvalues. Therefore, if this effect were the only contribution, it would be completely valid to calculate the losses without worrying about the exact wavelength, as the mode itself would not change if the wavelength were scanned to hit resonance. The second effect is the change in eigenvector composition due to wavelength, which occurs because, as wavelength is scanned, the basis modes intersecting the mirror change in waist, phase curvature, and phase relative to other modes. This component means that the cavity eigenmodes may change as the probe wavelength is scanned

\begin{figure}
\captionsetup{width=0.95\textwidth}
\centering\includegraphics[width=0.7\columnwidth]{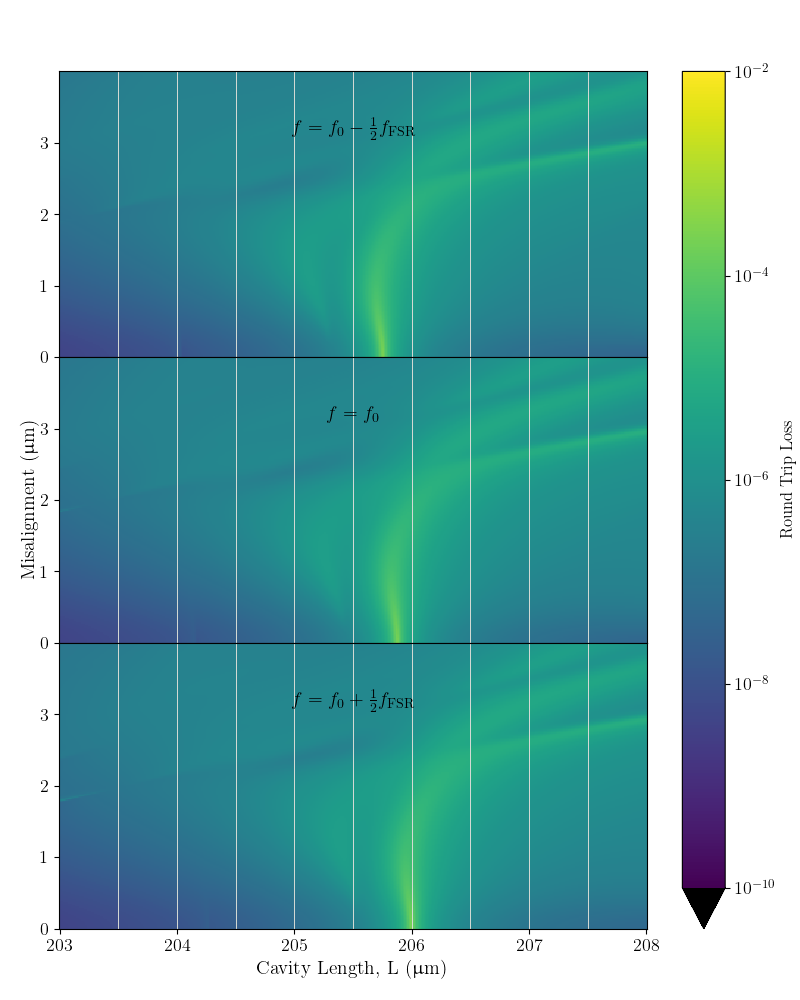}
\caption{Round-trip loss in a small region of the length-misalignment configuration space for a $2w_e=75~\upmu\mathrm{m}$ Gaussian mirror cavity interrogated at three wavelengths/frequencies that cover a free spectral range of tuning.}
\label{fig: fsr_range_data}
\end{figure}

Given the complexity of the calculations, it is difficult to argue analytically that tweaking the wavelength to probe the cavity resonances does not affect the numerical conclusions in the manuscript significantly. Therefore, this assumption was tested numerically by analysing data at three wavelengths. The first was a nominal central wavelength, and the other two had frequencies that differed from the central frequency on either side by half a free spectral range. This means that, at any point in the configuration space, the wavelength required to hit a resonance will lie within the bounds of the highest and lowest wavelength/frequency. The scans were taken over a sharp resonance feature, and are shown in Fig.~\ref{fig: fsr_range_data}. The wavelength change is shown to affect the positions of the resonances, but keep the structure mostly intact. This suggests that calculating losses for a single wavelength is a valid approach to investigate the qualitative behaviours seen in the loss maps of these systems, but that comparisons of the exact positions of lossy resonances to experiment should account for the wavelength scanning. Additionally, for shorter cavities with larger free spectral ranges, the wavelength scans required to hit resonances will become more extensive, and below a certain length it will become important to consider scanning effects for even the qualitative conclusions.

\section{Algorithm and its performance}
\label{app: algorithm}
The procedure used to produce the data in the manuscript follows the numerical steps described in Hughes et al.~\cite{Hughes:23_1}. First, the mirror matrices for aligned configurations and translation matrices are calculated using techniques that avoid numerical integration. Secondly, the round trip matrices for all misalignments are calculated. Thirdly, the round trip matrices are diagonalised numerically, and finally the mode of interest is determined. The code to perform this procedure was written in Python and uses the libraries numpy, scipy, and mpmath to perform calculations, with matplotlib used to produce the figures \cite{Numpy:20, Scipy:20, Mpmath:13, Matplotlib:07}. We developed this code because we found that methods calculating the mirror matrix using numerical integration tended to be much slower. We have not, however, performed a comprehensive algorithmic comparison.

The calculations themselves were performed on the University of Oxford Advanced Research Computing (ARC) facility. http://dx.doi.org/10.5281/zenodo.22558. A typical dataset contained 800 cavity length samples and 400 misalignment samples. For the Gaussian mirror data, such calculations were run on 800 nodes, which each took roughly 24 hours to produce data for the 400 misalignment values over a basis size of 5000. The longest individual processes in the computation were the calculation of mirror matrices. However, due to the large number of translation samples, the biggest overall contribution to the run time was the diagonalisation of round trip matrices to calculate eigenmodes.



\end{document}